# Persistent large anisotropic magnetoresistance and insulator to metal transition in spin-orbit coupled antiferromagnets $Sr_2(Ir_{1-x}Ga_x)O_4$

Haowen Wang[1], Wei Wang[1], Ni Hu[2], Tianci Duan[2], Songliu Yuan[1], Shuai Dong[3], Chengliang Lu[1] [*], and Jun-Ming Liu[4,5]

[1]*School of Physics & Wuhan National High Magnetic Field Center, Huazhong University of Science and Technology, Wuhan 430074, China*

[2]*School of Science and Hubei Collaborative Innovation Center for High-Efficiency Utilization of Solar Energy, Hubei University of Technology, Wuhan 430068, China*

[3]*School of Physics, Southeast University, Nanjing 211189, China*

[4]*Laboratory of Solid State Microstructures and Innovation Center of Advanced Microstructures, Nanjing University, Nanjing 210093, China*

[5]*Institute for Advanced Materials, Hubei Normal University, Huangshi 435001, China*

[*] Email: cllu@hust.edu.cn

# Abstract

Antiferromagnetic (AFM) spintronics, where magneto-transport is governed by an antiferromagnet instead of a ferromagnet, opens fascinating new perspectives for both fundamental research and device technology, owing to their intrinsic appealing properties like rigidness to magnetic field, absence of stray field, and ultrafast spin dynamics. One of the urgent challenges, hindering the realization of the full potential of AFM spintronics, has been the performance gap between AFM metals and insulators. Here, we demonstrate the insulator-metal transition and persistently large anisotropic magnetoresistance (AMR) in single crystals $Sr_2(Ir_{1-x}Ga_x)O_4$ ($0 \leq x \leq 0.09$) which host the same basal-plane AFM lattice with strong spin-orbit coupling. The non-doped $Sr_2IrO_4$ shows the insulating transport with the AMR as big as ~16.8% at 50 K. The Ga substitution of Ir allows a gradual reduction of electrical resistivity, and a clear insulator-to-metal transition is identified in doped samples with $x$ above 0.05, while the AMR can still have ~1%, sizable in comparison with those in AFM metals reported so far. Our experiments reveal that all the samples have the similar fourfold AMR symmetry, which can be well understood in the scenario of magnetocrystalline anisotropy. It is suggested that the spin-orbit coupled antiferromagnets $Sr_2(Ir_{1-x}Ga_x)O_4$ are promising candidate materials for AFM spintronics, providing a rare opportunity to integrate the superior spintronic functionalities of AFM metals and insulators.

**Keywords:** antiferromagnetic spintronics, anisotropic magnetoresistance, spin-orbit coupling

**Introduction**

Antiferromagnets have been garnering increasing interest in the spintronics community, owing to their attractive properties and their technological promise in leading toward efficient memory devices [1-8]. The appealing features, such as rigidity to external magnetic field, absence of stray field, and ultrafast spin dynamics, make the antiferromagnetic (AFM) materials particularly favorable for ultrahigh density and high speed spintronics [7]. A major breakthrough in this emerging AFM spintronics field is the observation of anisotropic magnetoresistance (AMR) in several antiferromagnets recently [1,3,4,6,8,9], which confirmed the viability of proposal for utilizing antiferromagnets in spintronic devices [10]. When the staggered spin orders in antiferromagnets are manipulated by various approaches, the equilibrium relativistic electronic structure is expected to be modified. As a consequence, the AFM orders can be electrically readout by AMR effect which is the magnetotransport counterpart of the relativistic energy anisotropy.

The studies of AFM-based AMR (AFM-AMR) have well illustrated the leading role of strong spin-orbit coupling (SOC) in manipulation and detection of the AFM orders. This has stimulated the search for new materials combining the Néel orders and strong SOC effect. A straightforward and efficient way that has been utilized to involve strong SOC is to alloy $3d$ magnetic species with heavier elements. For instance, a spin-valve-like magnetoresistance (MR), as large as 160%, was observed in a MnIr-based tunneling junctions [1], and room temperature ($T$) memory was revealed in a simple resistor made of AFM FeRh [4]. Recently, the discovery of a novel $J_{\text{eff}}$=1/2 state in several $5d$ iridates has posed a fundamentally distinct scheme to build up a spin-orbit coupled AFM lattice where the SOC and magnetic moment have the same source, i.e. the Ir ions [11-15]. In fact, cases of large SOC involved in building up the electronic and magnetic structures have been widely seen in $4d/5d$ magnetic materials [16,17]. Because of the pronounced SOC (scaled by factor $\lambda \sim 0.5$ eV) at the Ir site in iridates, the $t_{2g}$ band of Ir $5d$ state is split into half-filled $J_{\text{eff}}$ =1/2 and fully occupied $J_{\text{eff}}$ =3/2 bands. Further, the $J_{\text{eff}}$ =1/2 band can be split by a moderate Hubbard repulsion, leading to a Mott transition [11]. Interestingly, the $J_{\text{eff}}$ =1/2 moment, entangling both spin and orbital momenta, was found to be a very efficient ingredient for AFM-spintronics. In the prototypical $J_{\text{eff}}$ =1/2 AFM insulator $Sr_2IrO_4$, giant relativistic energy anisotropy was predicted upon the AFM spin-axis rotation [3], and a remarkable AMR of ~14% was experimentally evidenced [5]. Although fascinating functionalities have been identified in the AFM metallic alloys and insulating oxides separately, it remains a significant challenge to integrate these superior properties properly, which hinders the realization of the full potential of AFM spintronics. For instance, the AFM metals are suitable for electrical information writing, while the AFM insulators are more compatible with transistors.

Difficulties for closing the performance gap between AFM metals and insulators lie mostly in finding appropriate materials in which the electric transport can be tuned over a broad range, i.e. from metallic to insulating, without breaking the inherent AFM lattice and especially the related spintronic functionalities, e.g. the AMR effect. This has actually been one of the main obstacles to the advancement of AFM spintronics. As compared with the AFM alloys showing robust metallicity, the $J_{eff}$ =1/2 Mott state in iridates, i.e. in $Sr_2IrO_4$, is found to be rather tunable, such as the insulator-metal transition induced by doing [18-21]. In particular, hidden orders like pseudogaps and Fermi arcs were experimentally identified in doped $Sr_2IrO_4$, in analogy to those found in the high temperature ($T$) cuprate superconductors [22-24].

To realize an AFM-AMR effect in $Sr_2IrO_4$ and its derivates, robust AFM lattice is strongly desired and has to be maintained in the doping induced metallic state. So far, the AFM-AMR has already been identified in the non-doped $Sr_2IrO_4$, which was understood as a consequence of magnetocrystalline anisotropy [3,5]. However, previous works focusing on the insulator-metal transition in doped $Sr_2IrO_4$ have commonly found that the basal-plane AFM lattice is prone to collapse upon the presence of metallic state [18-21]. This is intriguing indeed, since a signature character of $Sr_2IrO_4$ is the absence of anomaly in the electric transport corresponding to the AFM transition at $T_N$~240 K [19]. In fact, several recent theoretical calculations predicted the coexistence of AFM order with metallic state in $Sr_2IrO_4$ upon pure carrier doping [25-27]. To achieve this, nonmagnetic species doping should be more preferred, which can exclude possible additional magnetic perturbations bringing into the system. In addition, substituting Ir with nonmagnetic ions in $Sr_2IrO_4$ may further enhance the local Ir(Ga)-$p$/O-$d$ hybridization and thereby the structure distortion, which usually favors the AFM interaction. Indeed, Liu *et al.* theoretically revealed a vital role of the $p$-$d$ hybridization in tuning structure distortion in doped $Sr_2IrO_4$ [28]. Therefore, doping the novel $J_{eff}$ =1/2 Mott state with nonmagnetic species in $Sr_2IrO_4$, in comparison with those magnetic doping, provides a promising route to solve the urgent issue of the apparent incompatibility of robust AFM order with metallic transport.

In the present work, we demonstrate successive modulation of the electric transport from insulating state to metallic state in nonmagnetic-Ga doped $Sr_2IrO_4$ single crystals, while the robust AFM orders and persistently large crystalline AMR can be retained. For $Sr_2IrO_4$, the AMR is found to be as big as ~16.8% at $T$=50 K, while the doped samples show not only metallic transport but also the AMR larger than ~1%. All the samples show the same fourfold AMR symmetry, arising from the magnetocrystalline anisotropy.

**Experiments**

The $Sr_2(Ir_{1-x}Ga_x)O_4$ ($0 \leq x \leq 0.09$) single crystals were synthesized from off-stoichiometric quantities of $SrCl_2$, $Sr_2CO_3$, $IrO_2$, and $Ga_2O_3$ using the self-flux techniques. In order to determine the crystalline structure, select crystals were crashed thoroughly and then powder X-ray diffraction (XRD) measurements were performed at room temperature. The ionic oxidation state of Ir was determined by performing the X-ray photoelectron spectroscopy (XPS) measurements. Atomic force microscopy characterizations were performed to capture the surface morphology of the crystals after cleaving.

Electric transport measurements for all the samples were carried out using a four-probe method in a Quantum Design (QD) physical property measurement system equipped with a rotator module. With regard to the AMR measurements, exciting current $I$ was applied along the [001] direction, and the magnetic field $H$ was always rotated within the basal plane of the crystals. Therefore, the AMR refers to the ($I$, $H$) behaviors. Magnetization ($M$) as function of $T$ and $H$ were measured using a QD superconducting quantum interference device. During the $M(T)$ measurements in both the in-plane and out-of-plane geometries, the measuring field was fixed at $H$=0.1 T. All the $M(H)$ curves were measured after the zero-field cooled (ZFC) sequence.

**Results**

All the samples with different Ga-doping content $x$ show the pure phase with a tetragonal crystalline structure, evidenced by the measured XRD patterns presented in the Supplemental Material Figure S1. To check the crystal quality, sample with $x$=0.07 was cleaved, and then the surface morphology was mapped out through atomic force microscopy characterization, from which the layered structure can be clearly seen (Supplemental Material Figure S2). To obtain more details on the structure variation upon the Ga-doping, Rietveld profile refining of the XRD data was performed, and the refined details for the sample with $x$=0.05 are presented in Figure 1. The difference between the measured and refined spectra is small with reliability parameter $R_{wp}$=6.78%. For all the other samples, $R_{wp}$ values are at similar levels. The lattice parameters, including $a$ and $c$ values, and Ir-O-Ir bending angle $\varphi$ and IrO$_6$ octahedron rotation angle $\phi$ (= (180°-$\varphi$)/2) are shown in Fig. 1(b) and (c), respectively. Both the $a$ and $c$ values just show tiny variation upon the doing, as small as ~ 0.1% and ~ -0.2%, respectively, over the $x$-range from 0.0 to 0.09. This behavior is expected, because of the approximate ionic radius of Ga$^{3+}$ (0.620 Å) and Ir$^{4+}$ (0.625 Å) [29]. On the contrary, both the $\varphi$ and $\phi$ values show non-negligible variations upon the Ga-doping, e.g. $\varphi$ is enhanced by ~4° while $\phi$ is reduced by ~2° as $x$ is up to 0.05. As revealed by previous works, both the $\varphi$ and $\phi$ are important to the electronic and magnetic properties of Sr$_2$IrO$_4$ [30,31], as evidenced hereafter.

The Ga substituting for Ir in Sr$_2$IrO$_4$ is expected to introduce holes to the system, owing to the lower valance state of Ga$^{3+}$ than Ir$^{4+}$. This is confirmed by our XPS measurements and Hall effect measurements with data in the Supplemental Material Figure S3 and Figure S4, respectively. By fitting the XPS data with multiple valance state of Ir ions, the presence of Ir$^{5+}$ in the doped samples is identified, and the content of Ir$^{5+}$ is increased with increasing $x$. The charge accumulation surrounding Ir ions would promote the local Coulomb repulsion, and thus suppressing the Ir-$p$/O-$d$ hybridization and the structure distortion [28]. On the contrary, the introduction of nonmagnetic Ga ions is expected to enhance local $p$-$d$ hybridization and therefore the structure distortion, because of the empty $d$-orbit of Ga$^{3+}$. The two factors compete with each other and determine the structure distortion in the samples. It seems that as $x$<0.05 the sudden enhancement of hole carrier density in the samples lets the first factor stand out, giving rise to the enhanced Ir-O-Ir bond angle $\varphi$. Further increasing Ga content over 0.05 allows the second factor to be dominant, and as a consequence the structure distortion in the samples turns to be more serious, i.e. $\varphi$ is reduced at $x$>0.05.

Another effect of the charge-transfer process by the Ga-doping is the drastic reduction in resistivity ($\rho$), as shown in Fig. 2 (a), which is seen to be more than six orders of magnitude at low-$T$. The massive variation in electric transport in Sr$_2$(Ir$_{1-x}$Ga$_x$)O$_4$ crystals evidences the significant effect of hole-doping in tuning the $J_{eff}$=1/2 state. As anticipated, the non-doped Sr$_2$IrO$_4$ exhibits the insulating transport in the entire $T$-range, owing to the $J_{eff}$=1/2 Mott state [11].

Increasing $x$ causes the continuous suppression of the $\rho(T)$ curves, and finally triggers a metallic transport in samples with higher $x \geq 0.05$, evidenced by the positive slope of $\rho(T)$ curves shown in the inset of Fig. 2(a). More details of the slope change of $\rho(T)$ can be seen in Supplemental Material Figure S5. Certainly, one does observe the upturn in $\rho(T)$ in the low-$T$ range (<50 K), which is due to doping induced disorder effect commonly seen in doped $Sr_2IrO_4$ [18,32].

In accompanying with the tremendous modulation of electric transport and insulator-metal transition, it is critical to check whether the basal-plane AFM order remains robust against the Ga-doping or not. Fig. 2(b) and (c) show the measured magnetization data collected under $H$=0.1 T along the [100] and [001] directions, respectively, noting the very different scales of the $M$-axis for the two cases. All the samples uniformly show the sharp AFM transition, and the Néel temperature $T_N$ just presents the modest decrease from ~240 K to ~180 K upon increasing $x$ to 0.09. In particular, striking magnetic anisotropy with approximate $M_{ab}/M_c$~10 can be seen for all the samples. These features suggest that the basal-plane AFM configuration can be reserved for all the samples, which is supported by further magnetic characterizations shown below. The magnetic and transport properties are summarized in the phase diagram plotted in Fig. 2(d), where we will see the coexistence of AFM orders with different electronic states (insulator and metal).

In $Sr_2IrO_4$, a unique character of the basal-plane AFM order is the collective canting of Ir isospins due to the Dzyaloshinsky-Moriya (DM) interaction, which induces net magnetic moment in each $IrO_2$ plane [31,33,34]. As schematically shown in Fig. 3(a), within the $ab$-plane, the AFM-ordered isospins deviate from the $b$ axis with a certain angle $\alpha$. It was revealed that the canted Ir isospins rigidly track the $IrO_6$ rotation $\phi$ relative to the $c$ axis, leading to a locking effect of $\alpha \sim \phi$ [31]. This isospin canting leads to net magnetic moment along the $a$ axis with $M_a = M_{Ir} \cdot \sin\alpha \approx M_{Ir} \cdot \sin\phi$ in the $IrO_2$ plane [34]. In the ground state, $M_a$ is aligned antiferromagnetically along the $c$ axis without showing macroscopic magnetization. Upon a sufficient magnetic field $H$ higher than the critical field $H_{flip}$ for isospin flip within the $ab$-plane, an isospin-flip transition is triggered, and thus a weak ferromagnetic (FM) phase arises below $T_N$ [35,36]. This is responsible for the sudden $M$-enhancement at $T_N$ of the $M(T)$ curves (Fig. 2(b) and (c)). The induced weak FM phase can also be identified by well-shaped $M(H)$ curve with clear magnetic saturation, such as the one measured at $T$=10 K in $Sr_2IrO_4$ shown in Fig. 3(b). A closer look at the $M(H)$ curve reveals a step-like anomaly at ~ 0.2 T which is obviously the flipping field $H_{flip}$, as shown in Fig. 3(c), owing to the

$H$-driven isospin-flip transition. The complementary initial $M(H)$ curves after the ZFC sequence are shown in the Supplemental Material Figure S4.

Increasing $x$ to 0.09 doesn't cause apparent impact onto the magnetic saturation of all the samples, while the maximum magnetization $M_s$ derived at $H=5$ T decreases from ~0.093 to ~0.06 $\mu_B$/Ir, mainly ascribed to the hole-doping induced $Ir^{4+}$ to $Ir^{5+}$ conversion. Another visible effect of the Ga-doping is the appearance of FM-like hysteresis in samples with $x \geq 0.05$, as shown in Fig. 3(c). As revealed by previous neutron scattering experiments and theoretical calculations, hole-doping in $Sr_2IrO_4$ is expected to reverse $M_a$ alternatively, resembling the $H$ induced isospin-flip transition [37,38]. Nevertheless, the flipping of $M_a$ driven by the hole-doping is found to be irreversible, which well explains the FM-like hysteresis in the samples with relatively high doping content. Here it is noteworthy that the basal-plane AFM orders seen in the parent material $Sr_2IrO_4$ are expected to be preserved in all the samples, although the Ga-doping has modified the alignment of $M_a$.

As aforementioned, $M_a$ arises from the isospin canting, and is simply determined by the canting angle $\alpha$ (or the octahedral rotation angle $\phi$) and the Ir moment $M_{Ir}$ with an equation $M_a = M_{Ir} \cdot \sin\alpha \approx M_{Ir} \cdot \sin\phi$. This therefore provides a quantitative measure to verify the basal-plane AFM lattice in the doped samples. In Fig. 3(d), the calculated $M_a = M_{Ir} \cdot \sin\phi$ are in well agreement with the experimental values $M_s$ for all the samples. This demonstrates the basal-plane AFM lattice existing in all the samples. Here, $\phi$ is obtained through the structural refinement shown above, and $M_{Ir}$ is derived from the Curie-Weiss fitting of the magnetization data (Supplemental Material Figure S5).

To this stage, we have revealed continuous tuning of the electric transport from insulating state to metallic state and robust basal-plane AFM orders in Ga-doped $Sr_2IrO_4$ single crystals. This is also a quite unusual and emergent phenomenon, demonstrating that the nonmagnetic substitution of Ir in $Sr_2IrO_4$ could be an appreciated route to accommodate the coexistence of AFM order and metallic state. Then we turn our focus to track the spintronics functionality, i.e. the AMR effect, which is the main motivation of the present work, noting that a realization of AMR represents an important step towards the manipulation and detection of AFM orders, in analogy to the traditional FM spintronics. As shown in Fig. 4(a) and (b), all the samples with either insulating or metallic transport show the same fourfold AMR=$[R(\Phi) - R(0)]/R(0)$ symmetry, while the AMR amplitude is gradually reduced upon the Ga-doping. The device geometry of the AMR measurements is shown in Fig. 4(c), where the electric current is applied along the [001] direction, and $H$ is rotated within the basal plane during the measurements. The AMR amplitude as a function of $x$ derived at $T$=50 K and 90 K are summarized in Fig. 4(d). In $Sr_2IrO_4$, the AMR is found to be as huge as ~16.8% at $T$=50 K, in agreement with previous report [5]. Doping Ga at Ir-site up to $x$=0.05 reduces the AMR amplitude from 16.8% to 1%. For the samples with $x \geq 0.05$ where the metallic transport emerges, the AMR amplitude evolves with $x$ steadily at a level of ~1%, which is indeed among the largest values reported so far in metallic AFM materials. The same fourfold AMR symmetry can be generally seen below $T_N$ in all the samples, while their magnitude show gradual decrease with increasing $T$, shown in Fig. 4(d).

**Discussion**

We will now discuss the observed insulator-metal transition, robust basal-plane AFM orders, and persistently large fourfold AMR in $Sr_2(Ir_{1-x}Ga_x)O_4$ (0≤$x$≤0.09) single crystals. The insulator-metal transition has been frequently observed in electron or hole doped $Sr_2IrO_4$. Regarding the electron-doping in $Sr_2IrO_4$, the Mott gap collapses as a result of band-filling, which was identified through optical spectroscopy characterizations [39]. However, the hole-doping induced transition to metallic state is a little more complicated. First, the hole-doping can move down the chemical potential to or near the top of the lower Hubbard band [40]. Second, doping with 3$d$ or 4$d$ species at the Ir-site is expected to reduce the effective SOC, thus shrinking the band-gap in the $J_{eff}$=1/2 state [20]. In the present work, the Ga-doping at Ir-site in $Sr_2IrO_4$ fulfills the two factors simultaneously, i.e. introduces holes to the Ir-site and reduces the effective SOC. Therefore, the downshift of chemical potential assisted by the reduction of effective SOC should be the origin of the metallic transport in $Sr_2(Ir_{1-x}Ga_x)O_4$ crystals. In addition, our structural analysis reveals a small amount increase (~2%) of the Ir-O-Ir bond angle $\varphi$, which would contribute to the conduction as well, according to the density functional calculations [30].

Previous studies revealed that the basal-plane AFM lattice of $Sr_2IrO_4$ is easy to be broken upon the doping [15,18,20,21]. However, it is noted that magnetic species were generally used for the substitution in these works, which would certainly bring in additional magnetic disturbance along with the carrier doping. In a recent work, it was experimentally found that a tiny 3% of isovalent $Tb^{4+}$ substituting for $Ir^{4+}$ completely suppressed the long range AFM orders in $Sr_2IrO_4$ [41]. Upon Mn-doping in $Sr_2IrO_4$, the isospins were reordered from the basal-plane to the $c$-axis, although typical insulating transport behavior was seen simultaneously [15]. These works demonstrated the significant role of magnetic perturbation in tuning the AFM coupling in $Sr_2IrO_4$. According to the recent theoretical calculations, the basal-plane AFM lattice behaves indeed robust against pure carrier doping in $Sr_2IrO_4$ [25-27]. Therefore, it is physically reasonable to anticipate the preservation of the basal-plane AFM lattice in the nonmagnetic Ga-doped $Sr_2IrO_4$ where only the hole carrier density is enhanced.

Our structural analysis reveals clear variation in the Ir-O-Ir bond angle $\varphi$ and the octahedral rotation $\phi$ upon the Ga-doping, which may be related to the Ga-doping induced charge-transfer as discussed above. In particular, it is found that $\varphi$ decreases again as $x>0.05$, evidencing the enhancement of structure distortion in the samples. This is expected to stabilize the AFM orders. As shown in Supplemental Material Figure S6, $T_N$ evolves with $x$ in a more modest manner at $x>0.05$ indeed, as compared with the case at $x<0.05$. This is consistent with the structural analysis. In addition, topological defects accompanying with the Ga-doping will surely impact the Ir AFM network and thus contribute to the continuous decrease in $T_N$ within the entire doping range.

The key experimental observation of the present work is the large fourfold AMR effect in all samples hosting remarkably different transport properties. This has yet been reported in previous studies focusing on the AFM spintronics. The realization of AFM-AMR effect is crucial for the manipulation and detection of AFM orders. The above magnetization data reveal that the weak FM phase can be induced by either magnetic field or hole-doping. This provides a natural handle to manipulate the basal-plane AFM orders, and thus triggering the AMR effect in the samples. During the AMR measurements, $H$ and thereby $M$ is always perpendicular to the applied electric current, which means that the AMR is determined solely by the varying angle between $M$ and crystal axes, fulfilling the scheme of magnetocrystalline anisotropy. Previous neutron scattering and optical microscopy characterizations have demonstrated that the magnetic easy axis of $Sr_2IrO_4$ is along the crystal axis, i.e. the $b$-axis [31,34]. This should be the same case in all doped samples, since they have the same basal-plane AFM configuration as evidenced by our magnetic characterizations. As shown in Fig. 4(a) and (b), the AMR minima of all samples has a periodicity of $\pi/2$, and appear right at the crystal axes, matching well with the tetragonal symmetry of the crystals. This clearly demonstrates the magnetocrystalline anisotropy origin of the observed persistent large AMR in all samples.

It is known that the crystalline AMR is determined by the anisotropy of the density of state (DOS) and the corresponding position of the chemical potential. In the present work, the Ga-doping is revealed to modify the $J_{\text{eff}}=1/2$ state significantly. Therefore, it is expected to see different AMR magnitude in the samples with different Ga-content. Moreover, it was theoretically revealed that the DOS anisotropy of the AFM insulator $Sr_2IrO_4$ is far larger than that of AFM metals, i.e. MnIr and $Mn_2Au$ [3,10]. This indicates that the AFM insulators are more proper for obtaining sizable AMR effect, as compared with the AFM metals. In Fig. 4(c) and (d), the AMR magnitude of the samples showing insulating transport ($x<0.05$) is overall much larger than that of the samples with $x\geq0.05$ where metallic transport appears, which is in agreement with the theoretical revealing [3,10].

**Conclusion**

To conclude, we have reported the first observation of persistent large AMR effects and insulator-metal transition in the $J_{\text{eff}}=1/2$ antiferromagnets $Sr_2(Ir_{1-x}Ga_x)O_4$ ($0\leq x\leq0.09$). Our results have revealed that the Ga-doping to the $J_{\text{eff}}=1/2$ Mott state in $Sr_2IrO_4$ can continuously suppress the insulating transport, and eventually lead to a metallic behavior. Meanwhile, the basal-plane AFM orders exhibits robustness against the Ga-doping and well exists in all samples. Importantly, the fourfold AMR is generally evidenced in all the samples with remarkably different transport properties, which can be understood as a consequence of magnetocrystalline anisotropy. It is noted that the AMR in these samples with metallic transport can still be as large as ~1%, which is yet among the largest values reported in AFM metals. These findings suggest that the spin-orbit coupled antiferromagnets $Sr_2(Ir_{1-x}Ga_x)O_4$ are favorable candidates for exploiting emergent phenomena and functionalities within the concept of AFM spintroncis.

**Acknowledgements:** This work was supported by the National Nature Science Foundation of China (Grant Nos. 11774106, 11674055, 51721001, and 51431006), the National Key Research Projects of China (Grant No. 2016YFA0300101).

**Figure captions:**

Figure 1. (a) Rietveld refinement for the sample with $x$=0.05. (b) Evaluated lattice parameters $a$ and $c$, and (c) Ir-O-Ir bond angle $\varphi$ and IrO$_6$ rotation angle $\phi$ as a function of Ga content $x$.

Figure 2. (a) Measured resistivity as a function of temperature for the samples with $x$=0, 0.01, 0.03, 0.05, 0.07, 0.09. The inset highlights the metallic transport of the samples with $x \geq 0.05$. (b) and (c) show the temperature dependence of magnetization measured with $H$//$ab$-plane and $H$// $c$-axis for all samples, respectively. (d) Summarized magnetic and transport properties. The $T_N$ were obtained from the in-plane (magenta dots) and out-of-plane (green cross) magnetization data, respectively.

Figure 3. (a) Sketch of the magnetic structure of Sr$_2$IrO$_4$. (b) Magnetization as function of $H$ measured at $T$=10 K for all samples. (c) The $M(H)$ curves at low field range for samples with $x$=0, 0.01, 0.05. (d) Plotted $x$ dependence of magnetic moment $M_{Ir}$ obtained from Curie-Weiss fitting, $M_s$ derived from the $M(H)$ curves, and $M_{cal}$ calculated using the equation of $M_{cal}$= $M_{Ir}$•sin$\phi$.

Figure 4. (a) and (b) show measured AMR data at $T$=50 K and $H$=3 T for all samples. (c) AMR magnitude obtained at $T$=50 K (black squares) and 90 K (green dots) as a function of Ga content $x$. The inset shows the device geometry of the AMR measurements. (d) Temperature dependence of AMR magnitude for all samples.

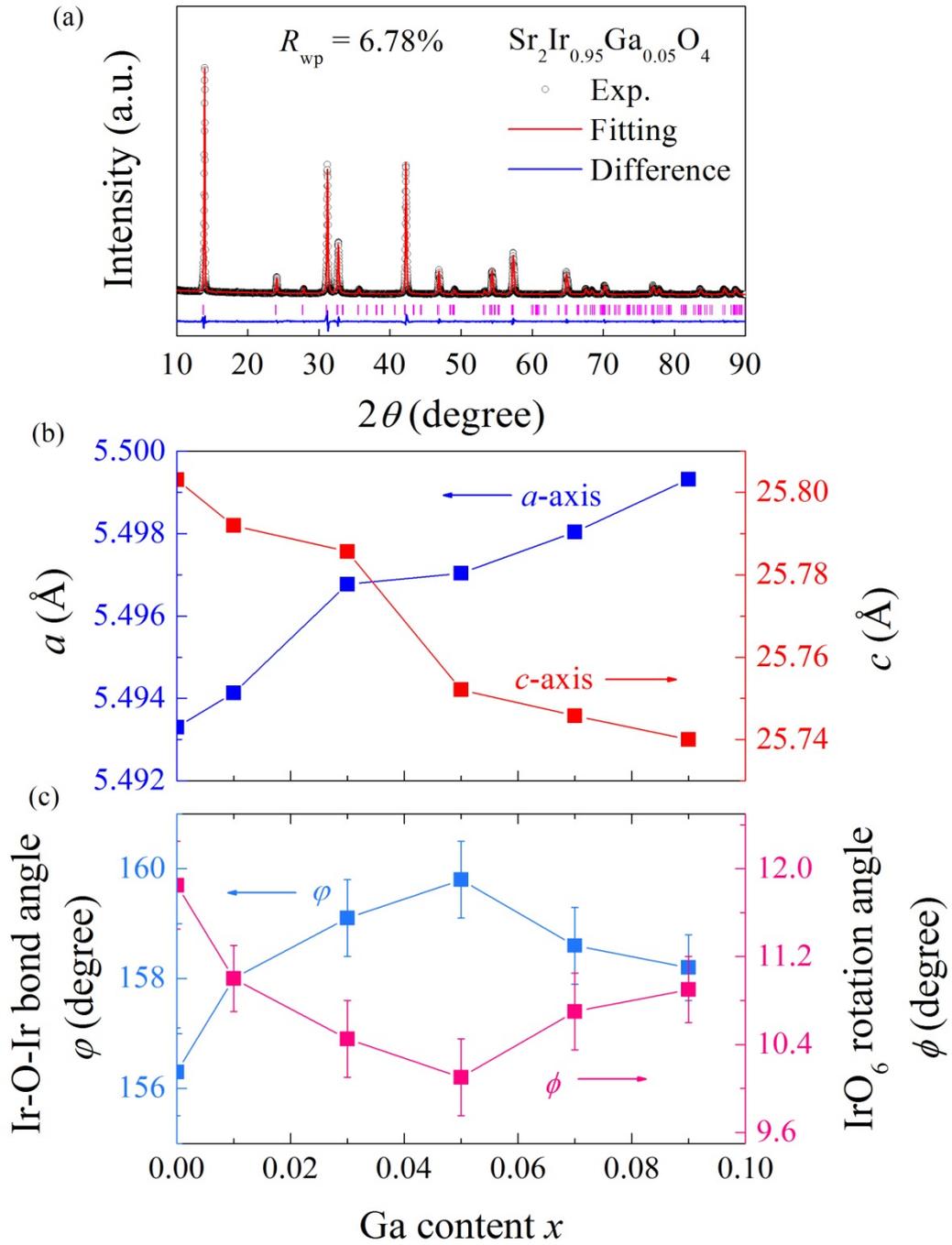

Figure 1.

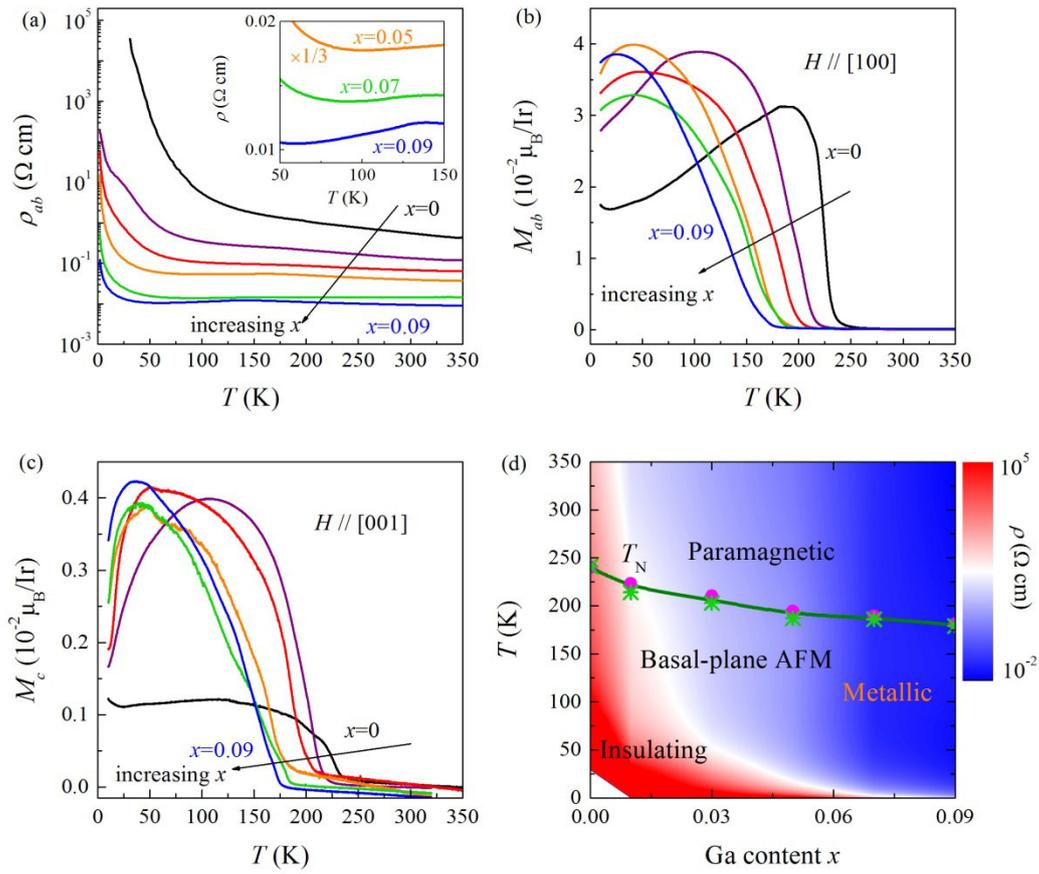

Figure 2.

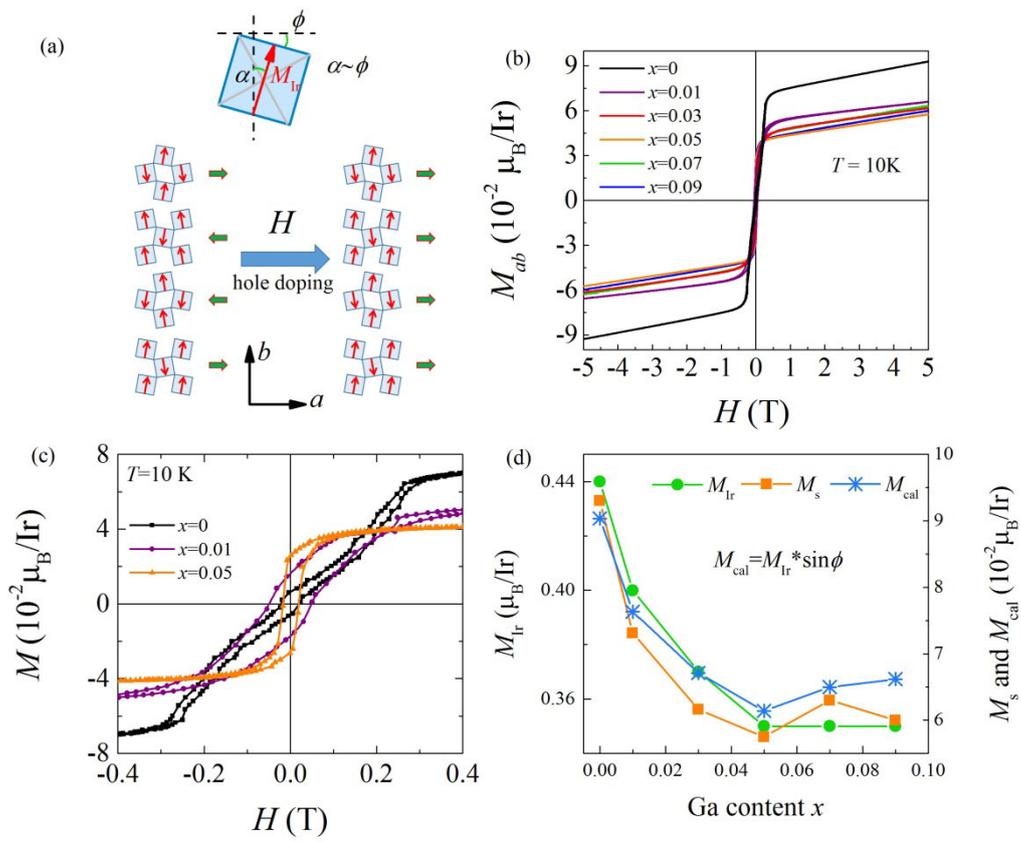

Figure 3.

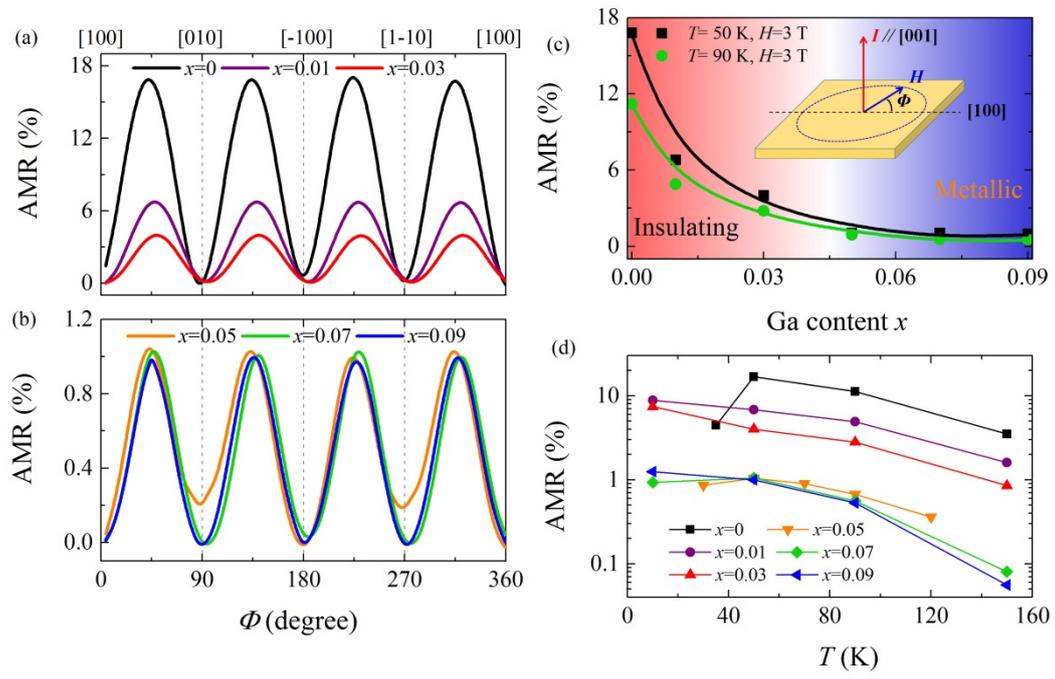

Figure 4.